# Self-Propulsion of a Metallic Superoleophobic Micro-Boat[1]

Albina Musin[a], Roman Grynyov[a], Mark Frenkel[a,b], Edward Bormashenko,[a,b*]

[a]*Ariel University, Natural Science Faculty, Physics Department, P.O.B. 3, 407000, Ariel, Israel*

[b]*Ariel University, Engineering Faculty, Chemical Engineering Department, P.O.B. 3, 40700, Ariel, Israel*

[*]Corresponding author: Edward Bormashenko

Ariel University, Natural Science Faculty, [a]Physics Department, [b]Chemical Engineering Department

P.O.B. 3, Ariel 40700, Israel

Phone: +972-3-906-6134; Fax: +972-3-906-6621

E-mail: edward@ariel.ac.il

Contains ESI (1 movie):

**Abstract**

The self-propulsion of a heavy, superoleophobic, metallic micro-boat carrying a droplet of various aqueous alcohol solutions as a fuel tank is reported. The micro-boat is driven by the solutocapillary Marangoni flow. The jump in the surface tension owing to the condensation of alcohols on the water surface was established experimentally. Maximal velocities of the self-propulsion were registered as high as 0.05 m/s. The maximal velocity of the center mass of the boat correlates with the maximal change in the surface tension, due to the condensation of alcohols. The mechanism of the self-locomotion is discussed. The phenomenological dynamic model describing the self-propulsion is reported.

---

## 1. Introduction

Autonomous locomotion of solid and liquid objects containing their own means of propulsion, called also self-propelling, driven by various physico-chemical (mainly interfacial) phenomena attracted considerable attention from researchers in the last decade [1–3]. Various mechanisms of self-propelling have been introduced, including the use of gradient surfaces [4–6], involving hot and cold Leidenfrost effects [7–11], soluto- and thermo-capillary Marangoni flows, and exploiting micro- and nano-structured surfaces [10–15]. Self-propulsion of micro-scaled objects and macroscopic bodies such as a camphor boat was investigated [16–17]. Self-propelling supported by liquid [10, 11, 18] and solid surfaces was reported.

The interest in self-propelling systems arises from numerous fundamental problems and applications, including mechanisms of the motion of bacteria and other microswimmers [20], lab-on-chip systems [15], drug delivery and microsurgery [21-22]. In our work we report self-propelling of a macroscopic superoleophobic boat, carrying aqueous solutions of various alcohols, driven by the solutocapillary Marangoni flow [23–25] (shown in Fig. 1A).

## 2. Experimental

For the manufacturing of the micro-boat aluminum plates with the thickness of 0.10 mm were used. The superoleophobic properties were conferred to aluminum plates by two-stage process described in detail in Ref. 26. At the first stage Al plates were immersed for 5 min in a 5%water solution of hydrochloric acid (HCl was supplied by Alfa Aesar). At the second stage dried micro-rough Al plates were immersed for 30 min in a solution of perfluorononanoic acid, 97% $C_9HF_{17}O_2$ (supplied by Alfa Aesar). Thus, micro-rough plates possessing pronounced superoleophobic properties were prepared [26]. The shape of the boat is shown in Fig. 1A-B.

The motion of the boat was registered by the epy-video imaging using a CASIO Digital Camera EX-FH20. After capturing the video, the movie was split into separate frames by the VirtualDub software. The videos were treated by the specially developed software, enabling the calculation of the speed of the boat.

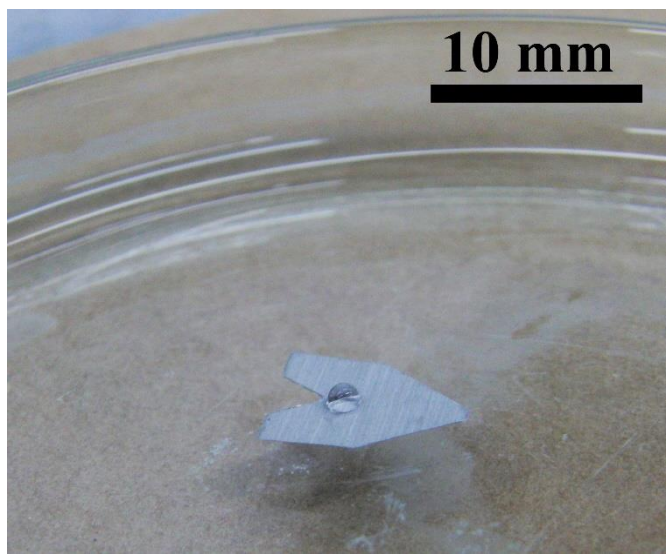

**Fig. 1A**. Superoleophobic metallic boat placed on the water surface carrying a 2.5 µl droplet of aqueous ethanol solution.

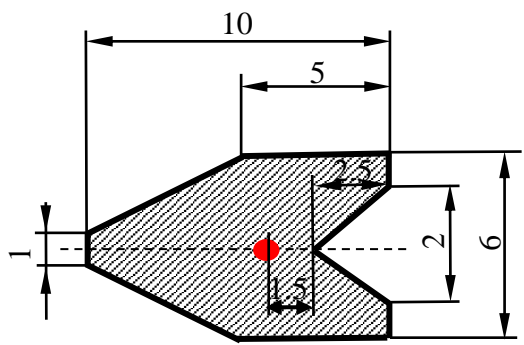

**Fig. 1B**. Geometric parameters of the boat (supplied in mm). The red spot depicts the deepening which fixed the location of a droplet.

Aqueous solutions of methanol (supplied by Tedia Co. Inc, USA, HPLC/SPECTRO grade), ethanol, propanol and butanol (supplied by Bio Lab Ltd.) Israel, AR grade), were used for creating Marangoni flows driving the boat. The general chemical formula of alcohols is $C_nH_{2n+1}OH$ ($n$ = 1 for methanol, 2 for ethanol, 3 for propanol, and 4 for butanol). The concentration of solution was varied within the range 5–40 wt%. The droplet of the solution with a volume of 2.5 µl was placed carefully with a precise micro-syringe on the surface of a floating boat as shown in Fig. 1A. The deepening shown with the red speckle in Fig. 1B enabled the fixation of the location of the droplet.

For the study of the impact exerted by alcohols evaporated from the droplet and condensed on the surface tension of the water surface following model experiments were carried out. Change in the surface tension of pendant water droplet was measured within different experimental scenarios. According to the first approach, a water droplet was suspended over alcohol surface, and the surface tension was measured *vs.* the distance between the alcohol surface and the bottom of the droplet (see Fig. 2). Under the second approach, water droplet was placed at some constant height above the alcohol surface, and the kinetics of the surface tension change was measured. Surface tension was measured with the pendant droplet method using the Ramé-Hart Advanced Goniometer Model 500-F1 at ambient conditions. The initial volume of water droplets was 5-6 µl.

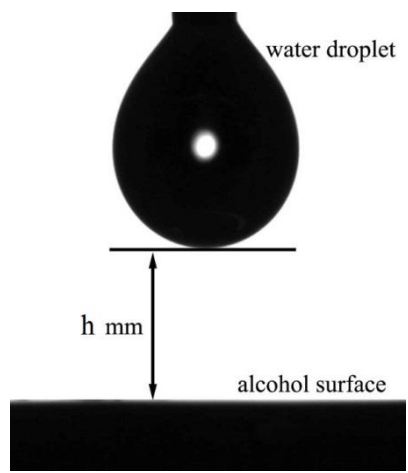

**Fig. 2**. The image of the pendant water placed at $h = 2$ mm height above the alcohol surface.

Figure 3 presents dependences of the surface tension on the separation of a pendant droplet from the alcohol surface. The surface tension was measured for the same water droplet when it moved down from 20 to 1 mm distance above the puddle of an alcohol. The dependence is very similar for all alcohols. The value of the surface tension decreases from 71 mJ/m$^2$ to the value slightly depending on the kind of the studied alcohols. Figure 4 presents changes of the surface tension of water droplets depending on the time of the pending above an alcohol surface. The water droplet was suspended above an alcohol puddle, and the surface tension was measured every minute at the rate 10 measurements at second, and the average was calculated.

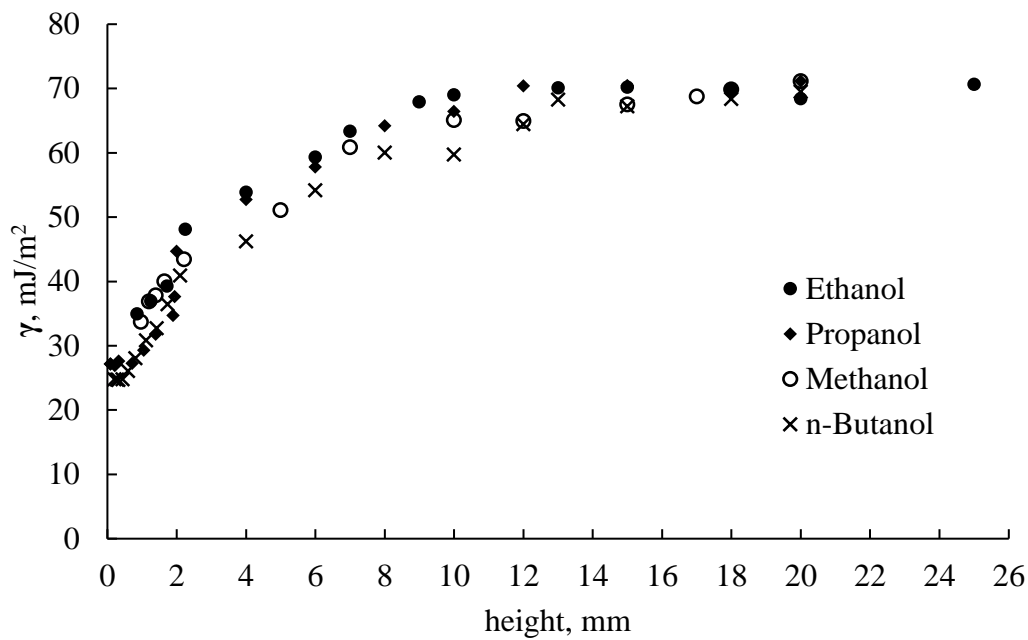

**Fig. 3**. The surface tension of the pendant water droplet vs. its separation from the alcohol surface for various alcohols.

It is recognized from the data supplied in Fig. 3 that the surface tension of the pendant water droplet, placed above the alcohols surfaces, approached to 25 mJ/m$^2$ with the decrease of the spatial separation $h$. Thus the maximal change in the surface tension of the droplet due to the condensation of alcohols vapors is estimated as approximately 45 mJ/m$^2$.

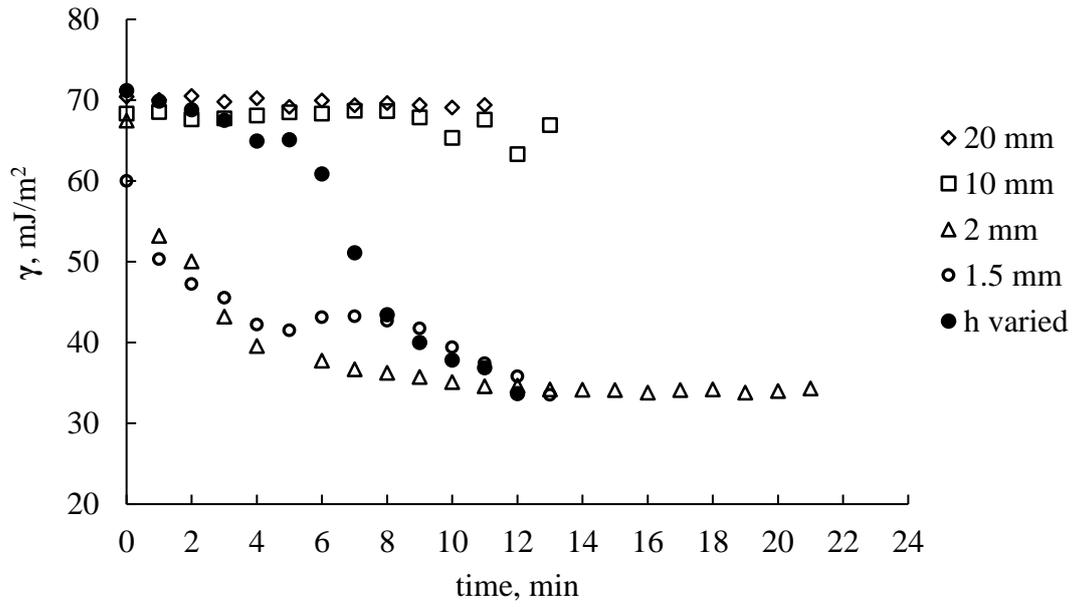

**A**

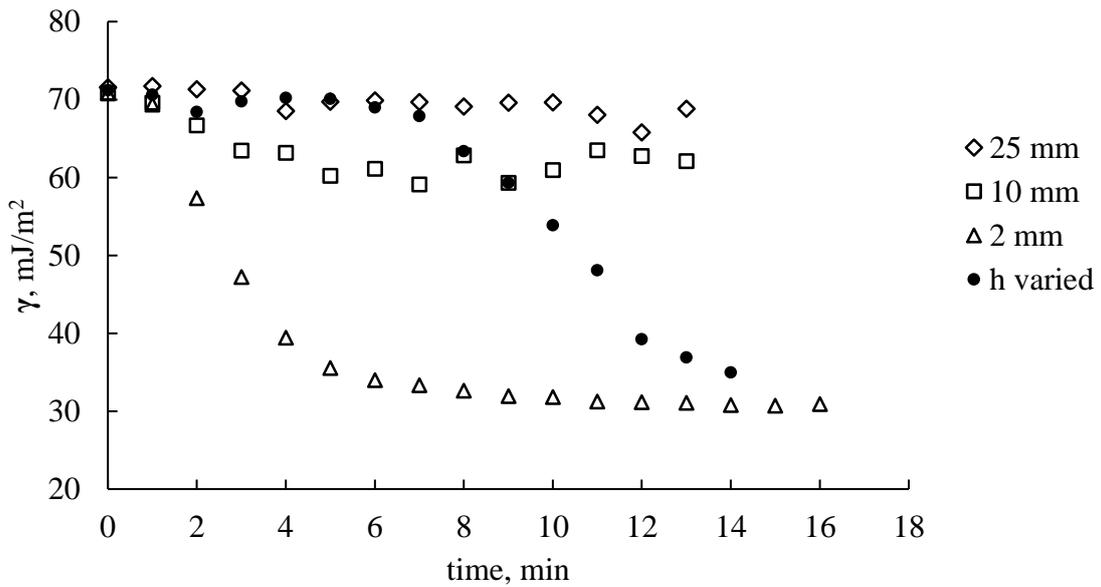

**B**

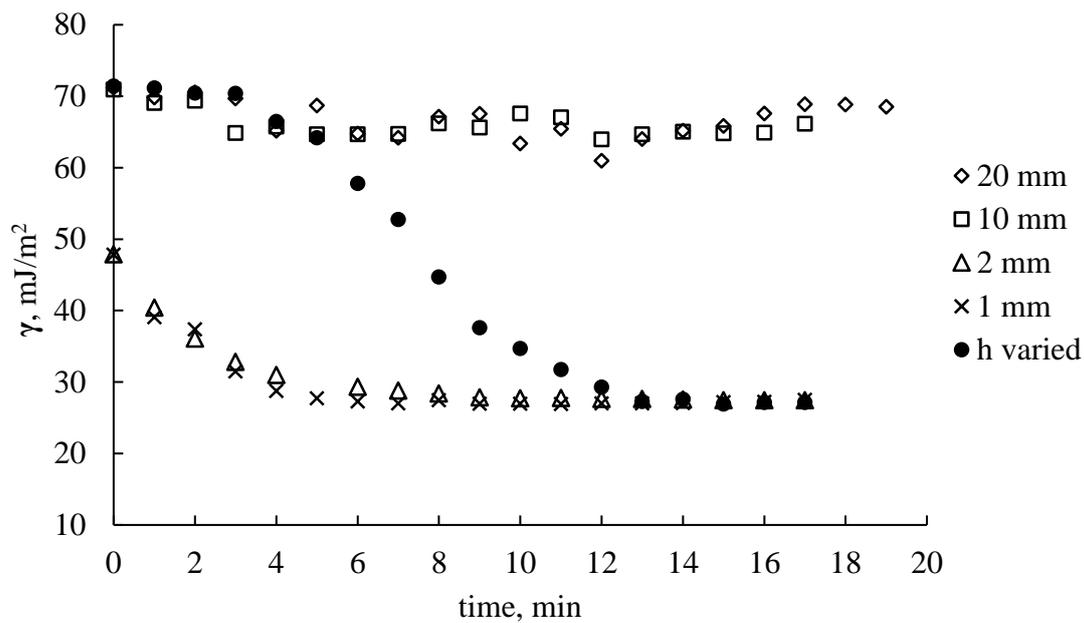

**C**

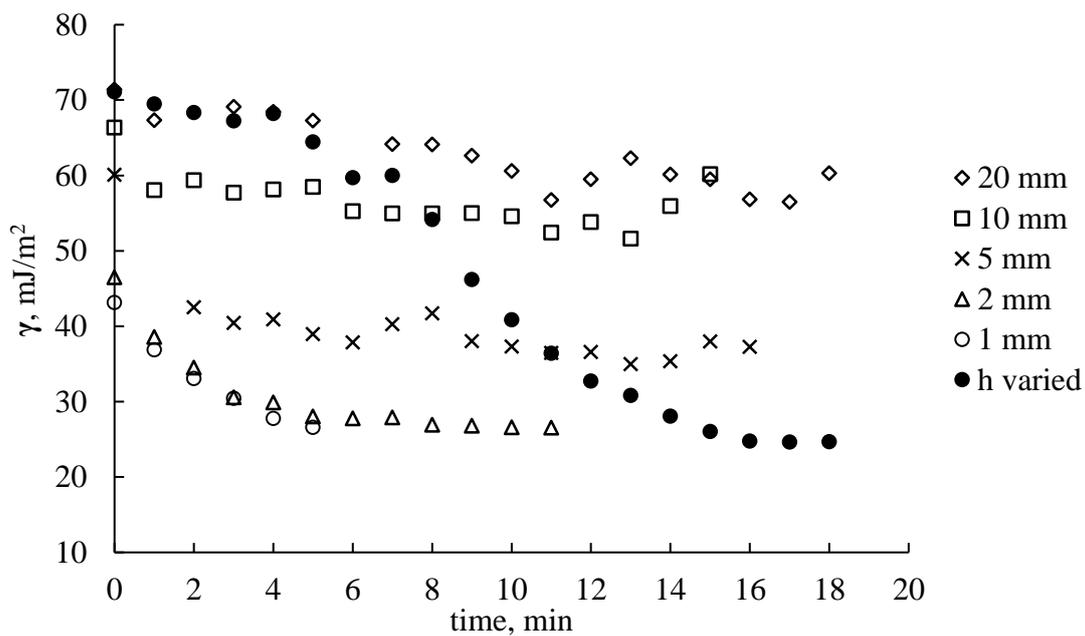

**D**

**Fig. 4.** The kinetics of the change of the surface tension of pendant water droplets placed at the fixed separation from: **A** methanol, **B** ethanol, **C** propanol, and **D** n-butanol.

Figure 4 A-D demonstrates that the surface tension of the pendant water droplet, placed at the constant separation from alcohol surfaces decreases from the initial value of 71 mJ/m$^2$ for pure water to a some saturation value which is very close to the surface tension of a pure alcohol (see Table 1). Closed circles at each graph correspond to the surface tension measured for the same water droplet when it moved down from 20 to 1 mm distance above the puddle of an alcohol on the same time scale according to the first approach. These graphs demonstrate that alcohol evaporates from the puddle and condenses on the surface of a pendant water droplet resulting in the decreasing of the surface tension.

**Table 1**. Final surface tensions of pendant water droplets exposed to various alcohol (C$_n$H$_{2n+1}$OH, $n = 1…4$) vapors.

|  | Methanol ($n = 1$) | Ethanol ($n = 2$) | Propanol ($n = 3$) | n-Butanol ($n = 4$) |
|---|---|---|---|---|
| Surface tension of a pure alcohol, $\gamma$, mJ/m$^2$, 25°C (*) | 22.1 | 22.0 | 20.9 | 25.0 |
| Measured surface tension of water droplet suspended above an alcohol, $\gamma$, ±0.2 mJ/m$^2$ | 34.0 | 30.8 | 27.4 | 26.6 |
| The difference between pure water surface tension (71.97 mJ/m$^2$, at 25°C) and surface tension of water droplet suspended above an alcohol, $\Delta\gamma$, ±0.2 mJ/m$^2$. | 38 | 41.2 | 44.6 | 45.4 |
| The maximal velocity of the center mass of the boat $v_{cm}^{max}$, carrying 2.5 µl droplet containing an aqueous solution ($c_0 = 5$ wt%) of the alcohol, ±0.001 m/s. | 0.003 | 0.005 | 0.018 | 0.024 |

The Petri dish in which the self-propulsion was observed was put in thermostatic conditions. The temperature at the water surface was controlled with the Therm-App infrared camera with an accuracy of ±0.1°.

## 3. Results and discussion

### 3.1. Experimental data representing the self-propulsion

Self-propulsion of the superhydrophobic boat was observed for all kinds of aqueous solutions of alcohols used in the investigation, when the initial concentration of the alcohols $c_0$ was confined within the range of 5–40% wt. The sequence of images demonstrating the self-propulsion is supplied in Fig. 5. It was impossible to place on the boat droplets with the larger concentrations of alcohols due to the Cassie-Wenzel transitions, which occurred on the superhydrophobic surface of the micro-boat [27]. The self-propulsion took place under isothermal conditions; the thermal camera did not recognize the temperature change at the water surface with an accuracy of 0.1°. Hence the effects to the thermal Marangoni flows are negligible.

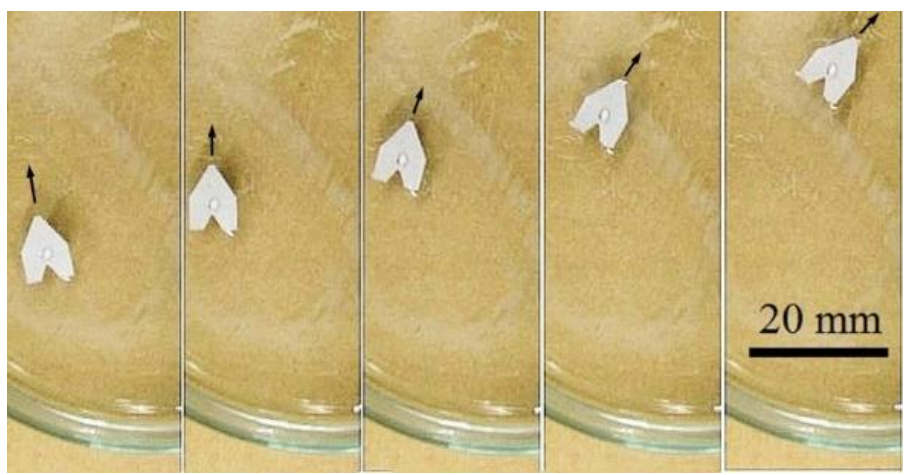

**Fig. 5**. The sequence of images demonstrating the self-propulsion of the superhydrophobic boat. The time separation between images is 2.66 s. The arrow shows the direction of the motion.

It is reasonable to suggest that the mechanism of the self-propulsion is similar to that discussed recently in Ref. 28. Jin *et al*. recently reported self-propulsion of a macroscopic boat driven by the diffusion of alcohols vapor (ranging from methanol to n-heptanol), through a superoleophobic nanofibrillar membrane, permeable to gases but repellent for liquid water and oil [28]. The evaporation of vapor decreased the original surface tension of the carrying liquid, allowing the self-propulsion of the boat [28].

In our experiments the spatial asymmetry of the surface tension, providing the Marangoni solutocapillary flow and consequent self-propulsion, was achieved by the special shape of the boat (depicted in Fig. 1A-B), enabling the condensation of the alcohol vapor mostly at the back of the boat, as illustrated in Fig. 6. The evaporation of various alcohols from the droplet created a surface tension gradient over the carrying liquid underneath the front and the rear parts of the boat and drove it forward. The mathematical model of the self-propulsion will be discussed below.

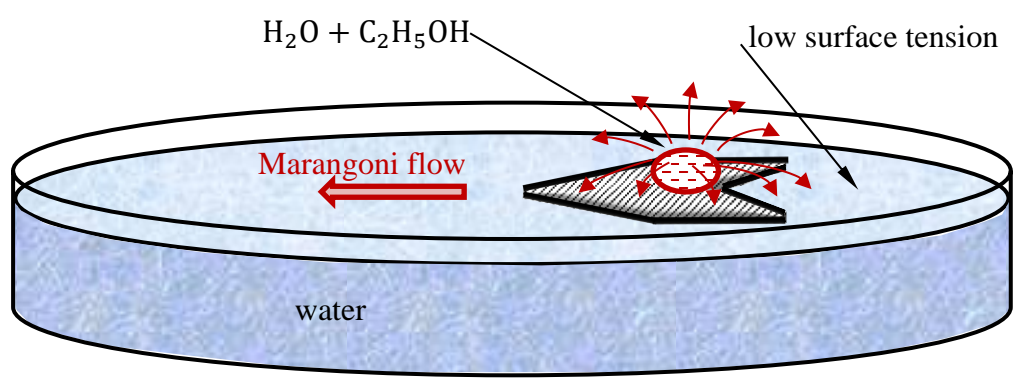

**Fig. 6**. Scheme illustrating the mechanism of self-propulsion of the metallic superoleophobic boat driven by the solutocapillary Marangoni flow. The boat is carrying on itself the droplet of an aqueous solution of alcohol. Alcohol evaporates, and due to the shape of the boat creates the area of the lower surface tension at the rear side of the boat. Red arrows show evaporation of alcohol.

The superoleophobicity [26, 29] of the boat provided it with the possibility to carry quasi-spherical droplets (the radius of the droplets was smaller than the capillary length) of water/alcohol solutions (serving as a fuel tank) which demonstrated markedly lower than pure water surface tensions (see the Experimental Section). It should be emphasized that the boat was made from aluminum, which is markedly heavier than the supporting water (the density of aluminum $\rho = 2.7$ g/cm$^3$). The ability of heavy objects to float was treated in detail recently in Ref. 30–31.

Two series of experiments were performed. In the first series 2.5 µl droplets of aqueous ethanol solutions of various concentrations (20–40 wt%) were carefully placed on the surface of the boat, as shown in Fig. 1A, and the self-propulsion of the boat was observed and registered. The results of these experiments are plotted in Fig. 7A–B. It is recognized from the graphs supplied in Fig. 7A–B that the self-propelling occurs in two distinct stages: accelerated motion which lasts *ca*. 10 s is followed by the decelerated motion which goes on *ca*. 100–150 s. The decelerated section of the motion is illustrated in more detail in Fig. 7B. It is seen from Fig. 7A–B that the maximal velocity of the self-propulsion grows, with the initial concentration of ethanol in the droplet $c_0$, whereas the deceleration is slightly dependent on the value of $c_0$.

In the second series of experiments 2.5 µl droplets of aqueous solutions of various alcohols with the same initial concentration ($c_0 = 5$ wt%) were placed on the boat. Figure 8 depicts the time dependence of the velocity of the center mass of the micro-boat $v_{cm}$ for the alcohols used in the investigation.

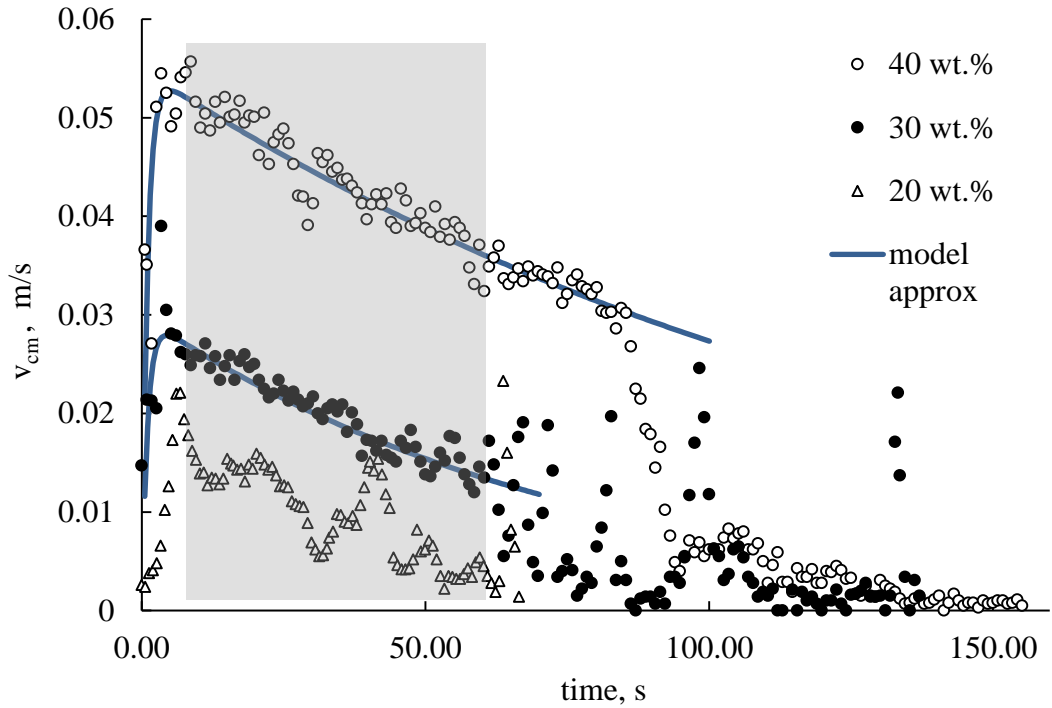

A

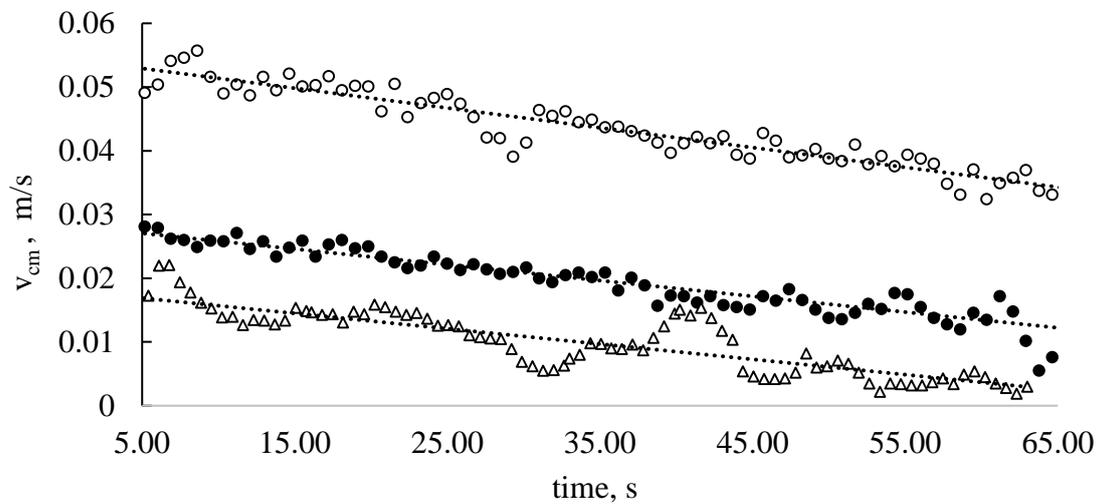

B

**Fig. 7. A** The time dependence of the velocity of center of mass of the boat $v_{cm}$ for water solutions of ethanol with various concentration $c_0$. The blue line represents fitting of the experimental data with Eq. (5).

**B** The decelerated section of the self-propulsion (marked by the gray rectangles in Fig. 7A).

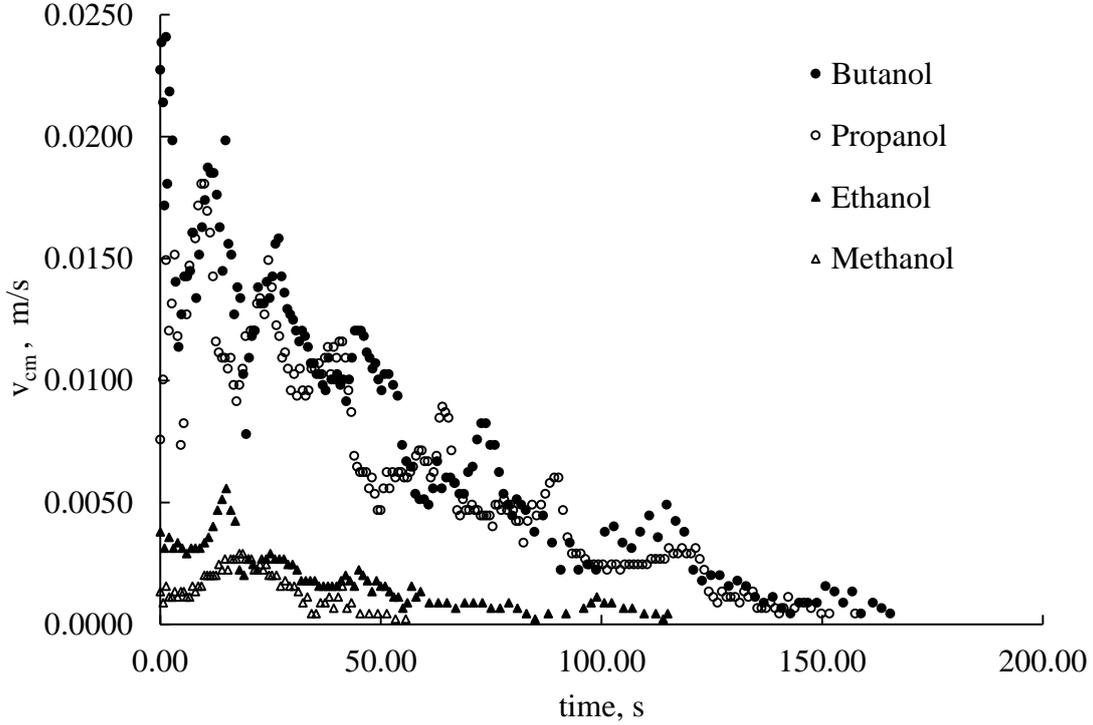

**Fig. 8**. The time dependence of the velocity of center of mass of the boat $v_{cm}$ for aqueous solutions of various alcohols with the initial concentration $c_0 = 5$ wt%

.

Again a relatively short stage of the accelerated self-propulsion (~10 s) is followed by the prolonged decelerated motion (~100–150 s). It should be emphasized that the maximal velocity of the center mass of the boat $v_{cm}^{max}$ unambiguously correlates with the maximal change in the surface tension, due to the condensation of alcohols, established experimentally with the pendant droplet method, as demonstrated in the Table 1. Thus it is reasonable to relate the effect of self-propulsion to the Marangoni soluto-capillary flow, resulting from the condensation of alcohols evaporated from a droplet on a water surface.

### 3.2. *The model describing the self-propulsion*

The equation describing the motion of the boat is:

$$m\frac{\mathrm{d}\vec{v}_{\text{cm}}}{\mathrm{d}t} = \vec{F}_{\text{fr}} + \alpha L^2 \nabla\gamma = -\chi L\eta\vec{v}_{\text{cm}} + \alpha L^2 \nabla\gamma \,, \tag{1}$$

where $m$, $L$ and $v_{\text{cm}}$ are the mass, characteristic dimension and velocity of the center mass of the boat correspondingly, $\gamma$ and $\eta$ are the surface tension and the viscosity of the supporting liquid correspondingly [32]; $\alpha$ and $\chi$ are the dimensionless coefficients, where $\chi$ depending on its shape, [32]. The accurate solution of Eq. (1) is a challenging task; assume "naively" that:

$$|\nabla\gamma| \cong \frac{|\Delta\gamma|_0}{L}\exp\left(-\frac{t}{\tau_{\text{ev}}}\right), \tag{2}$$

where $\tau_{\text{ev}} \cong 100\,\text{s}$ is the characteristic time of evaporation of alcohols in the solution, as established experimentally [32–33], $|\Delta\gamma|_0 \cong 45\,\text{mJ/m}^2$ is the initial jump in the surface tension due to the condensation of alcohols on the water surface (see the data supplied in Figs 3–4 and Table 1). Indeed, evaporation of alcohols decays with time, and we assumed that the decay is exponential in time (see Eq. (2)). Considering Eq. (2) enables rewriting of Eq. (1) for the one-dimensional self-propulsion as follows:

$$m\frac{\mathrm{d}v_{\text{cm}}}{\mathrm{d}t} = -\chi L\eta v_{\text{cm}} + \alpha L\Delta\gamma_0\exp\left(-\frac{t}{\tau_{\text{ev}}}\right). \tag{3}$$

It is convenient to re-shape Eq. (3) as given below:

$$\frac{\mathrm{d}v_{\text{cm}}}{\mathrm{d}t} + \frac{1}{\tau_{\text{fr}}}v_{\text{cm}} = \tilde{a}\exp\left(-\frac{t}{\tau_{\text{ev}}}\right), \tag{4}$$

where $\tau_{\text{fr}} \cong m/\chi L\eta$ is the characteristic time of viscous friction-based deceleration of the boat, $\tilde{a} \cong \alpha L\Delta\gamma_0/m$ is the constant with the dimension of the acceleration. The mass of the boat carrying a droplet is comprised of the mass of the boat itself ( $m_b = 0.018\,\text{g}$) and the mass of the droplet ($m_d = 0.002\,\text{g}$); thus, $m = m_b + m_d = 2 \times 10^{-5}\,\text{kg}$. For the sake of a very rough estimation, we assume: $L \cong 5 \times 10^{-3}\,\text{m}; \eta \cong 10^{-3}\,\text{Pa} \times \text{s}; \chi = 1$; this yields a rough estimation of the friction time: $\tau_{fr} \cong 5\,\text{s}$. The hierarchy of characteristic times also includes the characteristic time of diffusion of the ethanol vapor in air $\tau_{diff} \cong R^2 / D$ (where $R$ is the radius of the droplet and $D$ is the coefficient of diffusion of the ethanol vapor in air), and the characteristic time of the evaporation of a droplet $\tau_{ev}$ (which was established experimentally as *ca.* 2 min). Assuming $R \cong 10^{-3}\,\text{m}; D \cong 3 \times 10^{-5}\,\text{m}^2/\text{s}$ supplies for the characteristic time of diffusion the estimation $\tau_{diff} \cong 3 \times 10^{-2}\,\text{s}$; thus we conclude that in our experimental situation the inequality: $\tau_{diff} \ll \tau_{fr} \ll \tau_{ev}$ takes place. This hierarchy of characteristic times enables the self-propelling of the boat a carrying droplet-fuel tank, which continues for several seconds.

The true values of the parameters $\alpha$ and $\tilde{a}$ remain unknown, and we consider them as "free", fitting parameters. Solution of the differential equation (Eq. (4)) and considering the initial condition $v_{cm}(t = 0) = 0$ yields (consider that $\tau_{fr} \ll \tau_{ev}$):

$$v_{cm}(t) = \frac{\tilde{a}}{\tau_{fr}^{-1} - \tau_{ev}^{-1}} \left[ \exp\left(-\frac{t}{\tau_{ev}}\right) - \exp\left(-\frac{t}{\tau_{fr}}\right) \right] \cong \frac{\tilde{a}}{\tau_{fr}^{-1}} \left[ \exp\left(-\frac{t}{\tau_{ev}}\right) - \exp\left(-\frac{t}{\tau_{fr}}\right) \right], \quad (5)$$

Finally for the modulus of the acceleration of the center of mass of the boat $a_{cm} = \dfrac{d\,v_{cm}}{d\,t}$ we obtain:

$$a_{cm} \cong \frac{\tilde{a}}{\tau_{fr}^{-1}} \left[ \tau_{fr}^{-1} \exp\left(-\frac{t}{\tau_{fr}}\right) - \tau_{ev}^{-1} \exp\left(-\frac{t}{\tau_{ev}}\right) \right]. \qquad (6)$$

It is easily seen that $a_{cm} > 0$ and the motion is accelerated until $t < t^*$,

$$t^* = \tau_{\text{fr}} \ln \frac{\tau_{\text{ev}}}{\tau_{\text{fr}}} \approx 3\tau_{\text{fr}} \approx 15 \,\text{s} \,. \qquad (7)$$

It is convenient to introduce the dimensionless parameter $\varsigma = \tau_{\text{fr}} / \tau_{\text{ev}}$. The maximal velocity of the center of the boat is obtained after substitution of Eq. (7) into Eq. (5), and it equals:

$$v_{\text{cm}}^{\text{max}} = \frac{\tilde{a}}{\tau_{\text{fr}}^{-1}}(\varsigma^{\varsigma} - \varsigma) \cong \frac{\tilde{a}}{\tau_{\text{fr}}^{-1}}\varsigma^{\varsigma} \,. \qquad (8)$$

Considering $\tilde{a}$, $\tau_{\text{ev}}$ and $\tau_{\text{fr}}$ as fitting parameters we fitted the experimental data by the dependence supplied by Eq. (7). The resulting fitting curve is shown with the blue solid line in Fig. 7A. The values of fitting parameters calculated for various experimental conditions are summarized in Table 2. As it may be expected the values of the parameters $\tilde{a}$ and $\alpha$, representing in our model the driving force of the soluto-capillary Marangoni flow, grow with the concentration of ethanol in a droplet.

**Table 2.** Values of fitting parameters $\tilde{a}$, $\tau_{\text{ev}}$ and $\tau_{\text{fr}}$ extracted from the experimental data

| Ethanol, concentration, wt% | $\tilde{a}$ , m/s$^2$ | $\alpha$ | $\tau_{ev}$, s | $\tau_{fr}$, s |
|---|---|---|---|---|
| 30 | 0.030±0.001 | ~2.87×10$^{-6}$ | 75±6 | 0.93±0.18 |
| 40 | 0.055±0.001 | ~5.16×10$^{-6}$ | 143±10 | 0.95±0.1 |

As a "zero approximation" for the modulus of deceleration $a_{\text{cm}}$ we may deduce from Eq. (7) (for time spans $\tau_{\text{fr}} \ll t \ll \tau_{\text{ev}}$):

$$|a_{\text{cm}}| \cong \frac{\tilde{a}}{\tau_{\text{fr}}^{-1}\tau_{\text{ev}}} \,. \qquad (9)$$

It is expected from Eq. (9) that the values of deceleration for time spans $\tau_{fr} \ll t \ll \tau_{ev}$ will be of the same order of magnitude for various alcohols. This prediction is supported by the experimental data represented in Fig. 7B. Thus we came to conclusion that that the phenomenological dynamic of the self-propulsion casted by Eqs. (1)–(4) describes the phenomenon satisfactorily and successes to describe the essential features of the self-propulsion. It is noteworthy that the maximal velocities of the self-propulsion, listed in the Table 1, correlated with the maximal change in the surface tension, due to the condensation of alcohols, and in turn they correlated with the chain length of the molecule of alcohol. This observation contradicts, to the findings reported in Ref 28, where the inverse dependence of the velocity of self-propulsion on the chain length of the molecule of alcohol has been reported. This contradiction calls for the future investigation.

### 4. Conclusions

The self-propulsion of the heavy aluminum superoleophobic micro-boat is presented. The micro-boat is driven by the solutocapillary Marangoni flow, arising from the evaporation of various alcohols (namely: methanol, ethanol, propanol and butanol) from the "fuel tank", namely the 2.5 µl droplet containing the aqueous solution of alcohol. Use of the superoleophobic surfaces allowed placing on the micro-boat droplets containing aqueous solutions of alcohols with the concentration as high as 40 wt%. The special shape of the micro-boat provides the condensation of the ethanol on the supporting water surface mainly at the rear of the boat, thus the gradient of the surface tension is created under breaking of spatial symmetry [34]. This gradient, in turn, gives rise to the Marangoni flow, supplying to the boat maximal velocity, as high as *ca*. 0.05 m/s. The self-propulsion is stopped by the viscous drug.

The qualitative analysis of the self-propulsion is presented. The phenomenological model proposed in the paper describes satisfactorily a two-stage self-propelling motion of the boat, when the accelerated motion is followed by the decelerated displacement. Such kind of the Marangoni flow driven self-locomotion was also reported by other groups [35]. The introduced model is linear, thus the effects due to non-linearity of the problem, such as oscillation of velocity [36–37], clearly seen in Fig. 7A are lost and call for the improvement of the model.

The maximal velocities of the self-propulsion correlated with the experimentally established maximal changes in the surface tension, due to the condensation of alcohols, and in turn they correlated with the chain length of the molecule of alcohol. This observation contradicts, to the experimental observations discussed in Ref 28. In this research the self-propulsion of a nano-cellulose aerogel membrane, permeable to gases but repellent for liquids was reported. Vapors of various alcohols penetrated through the membrane and gave rise to the Marangoni flow [28]. The authors of Ref. 28 observed the inverse dependence of the velocity of self-propulsion of the membrane on the chain length of the alcohol molecule. This contradiction calls for the future research.

**Acknowledgements**

The authors are indebted to Mrs. Yelena Bormashenko for her kind help in preparing this manuscript. We also thankful to programmers from Kinematics & Computational Geometry Multidisciplinary Laboratory of Ariel University for their kind help in measurement software development.

Acknowledgement is made to the donors of the Israel Ministry of Absorption for the partial support of the scientific activity of Dr. Mark Frenkel.

**References**

[1] N. L. Abbott, O. D. Velev, Active particles propelled into researchers' focus, Current Opin. in Colloid and Interface Sci. 21 (2016) 1–3.

[2] J. Bico, D. Quere, Self-propelling slugs, J. Fluid Mech. 467 (2002) 101–127.

[3] Ph. T. Kühn, B. Santos de Miranda, P. van Rijn, Directed Autonomic Flow: Functional Motility Fluidics, Adv. Mater. 27 (2015) 7401–7406.

[4] S. Daniel, M. K. Chaudhury, J. C. Chen, Fast drop movements resulting from the phase change on a gradient surface, Science 291 (2001) 633–636.

[5] S. Daniel, S. Sircar, J. Gliem, M. K. Chaudhury, Ratcheting motion of liquid drops on gradient surfaces, Langmuir 20 (2004) 4085–4092.

[6] X.-P. Zheng, H.-P. Zhao, L.-T. Gao, J.-L. Liu, Sh.-W. Yu, X.-Q. Feng, Elasticity-driven droplet movement on a microbeam with gradient stiffness: a biomimetic self-propelling mechanism, J. Colloid & Interface Sci. 323 (2008) 133–140.

[7] J. Li, Y. Hou, Y. Liu, C.I. Hao, Mi. Li, M. K. Chaudhury, S. Yao, Z. Wang, Nature Physics 12 (2016) 606–612.

[8] R. L. Agapov, J. B. Boreyko, D. P. Briggs, B. R. Srijanto, S. T. Retterer, C. P. Collier, N. V. Lavrik, Asymmetric wettability of nanostructures directs Leidenfrost droplets, ACS Nano 8 (2014) 860–867.

[9] R. L. Agapov, J. B. Boreyko, D. P. Briggs, B. R. Srijanto, S. T. Retterer, C. P. Collier, N. V. Lavrik, Length scale of Leidenfrost ratchet switches droplet directionality, Nanoscale 6 (2014) 9293–9299.

[10] G. Lagubeau, M. Le Merrer, C. Clanet, D. Quéré, Leidenfrost on a ratchet, Nature Physics 7 (2011) 395–398.

[11] Ed. Bormashenko, Ye. Bormashenko, R. Grynyov, H. Aharoni, G. Whyman, B. P. Binks, Self-Propulsion of Liquid Marbles: Leidenfrost-like Levitation Driven by Marangoni Flow, J. Phys. Chem. C 119 (2015) 9910−9915.

[12] Ch. H. Ooi, A. van Nguyen, G. M. Evans, O. Gendelman, Ed. Bormashenko, N.-T. Nguyen, A floating self-propelling liquid marble containing aqueous ethanol solutions, RSC Adv. 5 (2015) 101006–101012.

[13] T. Ban, H. Nakata, Metal-Ion-Dependent Motion of Self-Propelled Droplets Due to the Marangoni Effect, J. Phys. Chem. B 119 (2015) 7100–7105.

[14] R. F. Ismagilov, A. Schwartz, N. Bowden, G. M. Whitesides, Autonomous Movement and Self Assembly, Angewandte Chemie 114 (2002) 674–676.

[15] W. F. Paxton, Sh. Sundararajan, Th. S. Mallouk, A. Sen, Chemical locomotion: Angewandte Chemie 45 (2006) 5420–5429.

[16] G. Zhao, M. Pumera, Marangoni self-propelled capsules in a maze: pollutants 'sense and act'in complex channel environments, Lab on a Chip 14 (2014) 2818-2823.

[17] S. Nakata, K. Matsuo, Characteristic self-motion of a camphor boat sensitive to ester vapor, Langmuir 21 (2005) 982–984.

[18] M. I. Kohira, Y. Hayashima, M. Nagayama, S. Nakata, Synchronized self-motion of two camphor boats, Langmuir 17 (2001) 7124–7129.

[19] G. P. Neitzel, P. Dell'Aversana, Noncoalescence and nonwetting behavior of liquids: Annual Rev. Fluid Mech. 34 (2002) 267–289.

[20] A. Snezhko, E. Ben Jacob, I. S. Aranson, Pulsating–gliding transition in the dynamics of levitating liquid nitrogen droplets: New J. Phys. 10 (2008) 043034.

[21] J. Elgeti, R. G Winkler, G. Gompper, Physics of microswimmers-single particle motion and collective behavior: a review: Reports Progress Physics 8 (2015) 056601.

[22] A. Ghosh, P. Fischer, Controlled propulsion of artificial magnetic nanostructured propellers: Nano Lett. 9 (2009) 2243–2245.

[23] J. Wang, Can man-made nanomachines compete with nature biomotors? ACS Nano 3 (2009) 4–9.

[24] A. A. Nepomnyashchy M. G. Velarde, P. Colinet, *Interfacial Phenomena and Convection* (Chapman & Hall/CRC Press, London/Boca Raton, 2002).

[25] P. G. de Gennes, F. Brochard-Wyart, D. Quéré, *Capillarity and Wetting Phenomena* (Springer, Berlin, 2003).

[26] A. Starostin, V. Valtsifer, V. Strelnikov, Ed. Bormashenko, R. Grynyov, Ye. Bormashenko, A. Gladkikh, Robust technique allowing the manufacture of superoleophobic (omniphobic) metallic surfaces: Adv. Eng. Mater. 16 (2014) 1127–1132.

[27] Ed. Bormashenko, Progress in understanding wetting transitions on rough surfaces, Adv. Colloid & Interface Sci. 222 (2015) 92–103.

[28] H. Jin, A. Marmur, O. Ikkala, R. H. A. Ras, Vapour-driven Marangoni propulsion: continuous, prolonged and tunable motion: Chemical Sci. 3 (2012) 2526.

[29] H. Bellanger, Th. Darmanin, E. T. de Givenchy, Fr. Guittard, Chemical and Physical Pathways for the Preparation of Superoleophobic Surfaces and Related Wetting Theories, Chem. Rev. 114 (2014) 2694–2716.

[30] D. Vella, Floating versus sinking: Annual Rev. Fluid Mech. 47 (2015) 115–135.

[31] Ed. Bormashenko, Surface tension supported floating of heavy objects: Why elongated bodies float better? J. Colloid & Interface Sci. 463 (2016) 8–12.

[32] C. H. Ooi, A. van Nguyen, G. M. Evans, O. Gendelman, E. Bormashenko, N.-Tr. Nguyen, A floating self-propelling liquid marble containing aqueous ethanol solutions: RSC Adv. 5 (2015) 101006–101012.

[33] C. H. Ooi, E. Bormashenko, A. V. Nguyen, G. M. Evans, Dz. V. Dao, N.-Tr. Nguyen, Evaporation of ethanol-water binary mixture sessile liquid marbles, Langmuir 32 (2016) 6097−6104.

[34] S. Michelin, E. Lauga, Autophoretic locomotion from geometric asymmetry, Eur. Phys. J. E 38 (2015) 7–22.

[35] G. Zhao, M. Pumera, Liquid−liquid interface motion of a capsule motor powered by the interlayer Marangoni effect, J. Phys. Chem. B 116, (2012) 10960−10963.

[36] M. Schmitt, H. Stark, Swimming active droplet: a theoretical analysis, EPL, 101 (2013) 44008.

[37] R. Sharma, S. T. Chang, O. D. Velev, Gel-based self-propelling particles get programmed to dance, Langmuir 28 (2012) 10128–10135.